\def\year{2022}\relax
\documentclass[letterpaper]{article} 
\usepackage{aaai22}  
\usepackage{times}  
\usepackage{helvet} 
\usepackage{courier}  
\usepackage[hyphens]{url}  
\usepackage{graphicx} 
\usepackage{arydshln}
\usepackage{latexsym}
\usepackage[switch]{lineno}
\usepackage{natbib}  
\usepackage{caption} 
\usepackage{algorithm}
\usepackage{algorithmic}
\usepackage{soul}
\usepackage{color, xcolor}
\usepackage{enumerate}
\usepackage{amssymb}
\usepackage{amsmath}
\usepackage{diagbox}
\usepackage{multirow}
\usepackage{makecell}
\usepackage{booktabs}
\usepackage{subfigure}
\usepackage{tabularx,ragged2e}
\usepackage{bbding}
\newcommand{\hlc}[2][yellow]{{\sethlcolor{#1}\hl{#2}}}
\definecolor{light_red}{rgb}{0.96, 0.76, 0.76}
\definecolor{light_green}{rgb}{0.7, 0.98, 0.87}
\definecolor{light_yellow}{rgb}{0.99, 0.97, 0.37}
\definecolor{dark_green}{rgb}{0.0, 0.5, 0.0}

\newcommand{\etal}{\textit{et al}. }

\newcommand{\eg}{\textit{e}.\textit{g}. }
\title{Tracing Text Provenance via Context-Aware Lexical Substitution}

\author{
    Xi Yang, 
    Jie Zhang\footnote{Corresponding authors.},
    Kejiang Chen,
    Weiming Zhang\footnotemark[1],\\
    Zehua Ma, 
    Feng Wang, 
    Nenghai Yu\\
}
\affiliations{
    University of Science and Technology of China\\
    \{yx9726@mail., zjzac@mail., chenkj@mail., zhangwm@, mzh045@mail., nishi@mail., ynh@\}ustc.edu.cn
}

\begin{document}
	\maketitle
	
	\begin{abstract}
		Text content created by humans or language models is often stolen or misused by adversaries. Tracing text provenance can help claim the ownership of text content or identify the malicious users who distribute misleading content like machine-generated fake news. There have been some attempts to achieve this, mainly based on watermarking techniques. Specifically, traditional text watermarking methods embed watermarks by slightly altering text format like line spacing and font, which, however, are fragile to cross-media transmissions like OCR. Considering this, natural language watermarking methods represent watermarks by replacing words in original sentences with synonyms from handcrafted lexical resources (e.g., WordNet), but they do not consider the substitution’s impact on the overall sentence's meaning.
		Recently, a transformer-based network was proposed to embed watermarks by modifying the unobtrusive words (e.g., function words), which also impair the sentence's logical and semantic coherence. Besides, one well-trained network fails on other different types of text content.
		
		To address the limitations mentioned above, we propose a natural language watermarking scheme based on context-aware lexical substitution (LS). Specifically, we employ BERT to suggest LS candidates by inferring the semantic relatedness between the candidates and the original sentence.
		Based on this, a selection strategy in terms of synchronicity and substitutability is further designed to test whether a word is exactly suitable for carrying the watermark signal. 
		Extensive experiments demonstrate that, under both objective and subjective metrics, our watermarking scheme can well preserve the semantic integrity of original sentences and has a better transferability than existing methods.
		Besides, the proposed LS approach outperforms the state-of-the-art approach on the Stanford Word Substitution Benchmark.
		
	\end{abstract}
	
	\section{Introduction}
	Tracing the provenance of text content is an important but still under-exploited issue in forensics. With readily available smart devices, adversaries can easily copy and distribute text content created by humans or language models, leading to undesirable consequences. For example, the leakage of confidential documents like unpublished literary works, commercial secrets, and government documents can often cause significant losses to individuals and society. Besides, powered by the advances of large-scale pre-trained language models like GPT-3 \cite{GPT-3}, natural language generation has made remarkable progress in generating fluent and realistic text. The adversaries can leverage these models to automatically generate misleading content like fake news \cite{fakenews} that look authentic and fool humans; or profit by plagiarising machine-generated valuable content such as financial reports \cite{financial_report}. 
	
	
	
	Watermarking is one of the techniques to solve the above issues, which has demonstrated its remarkable capabilities for protecting images \cite{hidden,stegastamp} and image processing networks \cite{model_watermarking}. However, it is more challenging to embed watermarks with imperceptible perturbations on text due to its inherent discrete nature.
	Traditional text watermarking schemes embed watermarks by slightly altering the image features like text format \cite{Brassil,unicode} and fonts \cite{fontcode,Qi_template}, which are fragile to cross-media transmissions like OCR. 
	Considering this, natural language watermarking (NLW) schemes choose to manipulate the semantics of text, which are inherently robust in the OCR-style transmissions. Most NLW works \cite{Topkara1,xiang} design a set of complex linguistic rules to substitute words with their synonyms chosen from handcrafted lexical resources like WordNet \cite{wordnet},
	but they fail to consider the substitution's impact on the overall meaning of the sentences. Moreover, it is time-consuming to build specific lexical dictionaries for different types of text content and the static dictionaries are not feasible for some linguistic phenomenons like polysemy. 
	
	Recently, an end-to-end transformer-based text watermarking network \cite{awt} was proposed to replace the unobtrusive words (e.g., articles, prepositions, conjunctions) in the input sentence with other inconspicuous words or symbols, which can guarantee the visual consistency between the watermarked text and the original text. Nevertheless, such replacements still impair the logical and semantic coherence of the sentences, because these selected words often represent specific semantic or syntactic information by forming phrases with their adjacent words. Besides, their dataset-specific framework has poor transferability on text content with other different writing styles.

	To address those limitations mentioned above, we propose a new context-aware lexical substitution (LS) approach and leverage it to build our watermarking scheme. Specifically, to avoid the dependence on static lexical resources and instead generate LS candidates for a target word directly based on its context, we explore the ``masked language model'' (MLM) pre-training objective of BERT to automatically generate LS candidates for a target word. Moreover, since the MLM-based generation only considers the probabilistic semantic similarity (SS) between the candidates and the target word, it is possible that two words in the candidates express opposite or unrelated meanings, such as `love' and `hate' in the masked sentence ``I \texttt{[MASK]} you''. 
	So we further introduce another BERT model to inference the semantic relatedness (SR) between the candidates and the original sentence, and then filter out the words that cannot maintain the original meanings. In this way, we can generate LS candidates by considering the overall sentence's meaning. That is, when the context changes, the candidates generated for the same word will change accordingly.
	
	
	However, this context-awareness poses a challenge for watermark embedding and extraction. Specifically, the candidates obtained from the original sentences will be different from those obtained from the watermarked sentences because it is inevitable to substitute some words in the original sentences to embed information.
	The challenge is, to achieve a successful message encoding and decoding, we must guarantee the candidates obtained from the original sentences and watermarked sentences are identical.
	Therefore, we design an LS-based sequence incremental watermarking scheme with a selection strategy in terms of synchronicity and substitutability, which enables the embedding and extraction sides can locate the same words and generate identical candidates for message encoding and decoding.
	
	In summary, our contributions are three-fold:
	\begin{itemize}
		\item We introduce the inference-based semantic relatedness into lexical substitution (LS) for guiding the candidates' generation. The proposed LS approach outperforms the state-of-the-art method on the Stanford Word Substitution Benchmark. It can be helpful in many NLP tasks like data augmentation and paraphrase generation.
		
		\item Based on the proposed LS approach, we design a sequence incremental watermarking scheme that can well preserve the meaning of the original text. And more than 80\% of the substituted original words can be recovered after watermark extraction. Besides, compared to existing methods, it requires no effort to design lexical resources or train networks and has a better transferability on different writing styles of text. 
		
		\item To our best knowledge, this is the first attempt to introduce a large-scale pre-trained language model for protecting text content created by humans or language models. We hope it can shed some light on this field and inspire more great works.
	\end{itemize}

	
	
	
	\section{Related Work}
	\paragraph{Natural Language Watermarking.} 
	Natural language watermarking (NLW) methods aim to embed watermarks by manipulating the semantics of sentences. 
	Existing works mainly construct static synonym dictionaries from WordNet and embed watermarks by synonym substitutions \cite{Topkara1}. Hao \etal \cite{xiang} introduced the word frequency ranking when choosing the synonyms to make the watermarked sentences look more natural. These methods have two limitations: (1) They fail to consider the substitution’s influence on the global semantics of the text, as some words can express different meanings in different contexts. (2) Depending on the type of text (news, novels, reviews, etc.), a specific synonym dictionary needs to be designed, which requires the participation of linguistic experts and is time-consuming.
	Recently, AWT \cite{awt} was proposed to using a transformer-based encoder-decoder network, trained on a specific dataset, to embed information in unobtrusive words with a given context. However, unobtrusive words such as articles, prepositions, and conjunctions often form common phrases with their adjacent words to express specific grammatical or semantic information. Therefore, although they are visually unobtrusive, the modified phrases may become incoherent.
	
	\paragraph{BERT-based Lexical Substitution.} The early studies \cite{WN-LS,SemEval2007,embedding} on lexical substitution also generate substitute candidates by finding synonyms from static lexical resources, which have the same limitations as the early NLW methods. Recently, it is demonstrated that BERT can predict the vocabulary probability distribution of a masked target word conditioned on its bi-directional contexts. Motivated by this, BERT-LS \cite{BERT-LS} was proposed and achieved the state-of-the-art results. In detail, it applies random dropout to the target word’s embedding for partially masking the word, allowing BERT to take balanced consideration of the target word’s semantics and contexts when generating substitute candidates. However, this method still searches for the semantic similar candidates in the word embedding space without considering the semantic relatedness. Besides, it cannot be used for NLW because the random dropout cannot guarantee that the generated candidates in the original and watermarked sentence are identical. But it still inspires us to leverage BERT for designing an LS-based watermarking scheme, which can further consider the semantic relatedness and does not rely on any static lexical resources or network training. 
	
	\section{Method}\label{Method}
	In this section, we will elaborate the proposed lexical substitution approach and leverage it to build the watermarking scheme. Before that, a brief description of the BERT model will be introduced.
	\subsection{Recap of the BERT Model}\label{BERT} 
	BERT is trained by two objectives: masked language modeling (MLM) and next sentence prediction (NSP). In the MLM-based training, a random token in the input sentence is replaced with the mask token \texttt{[MASK]}. Let $S = \{t_1, t_2,..., t_N\}$ represents the input sentence consisting of a set of tokens. As explained in \cite{BERTHA,LSim}, the MLM training is equivalent to optimizing the joint probability distribution:
	\begin{linenomath}
	\begin{equation}\label{joint_probability}
	\log P(S|\theta)=\frac{1}{Z(\theta)} \sum_{i=1}^{N} \log \phi_{i}(S|\theta),
	\end{equation}
	\end{linenomath}
	where $\phi_{i}(S|\theta)$ is the potential function for the $i$-th token with parameters $\theta$, $Z$ is the partition function. And the log-potential term is defined as:
	\begin{linenomath}
	\begin{equation}\label{log_potential}
	\log \phi_{i}(S|\theta)=t_{i}^{T} f_{i}\left(S_{\backslash i}| \theta\right),
	\end{equation}
	\end{linenomath}
	where $t_{i}^{T}$ is the one-hot vector of the $i$-th token. $S_{\backslash i}=\{t_1,...,t_{i-1},\texttt{[MASK]},t_{i+1},..., t_N\}$ and $f_{i}\left(S_{\backslash i}| \theta\right)$ is the output of the final hidden state of BERT corresponding to the $i$-th token for input $S_{\backslash i}$. 
	
	In the NSP-based training, two sentences are concatenated with a separator token \texttt{[SEP]}. And a classification token \texttt{[CLS]} will be added as the head of the input. A classifier is appended upon the final hidden state corresponding to the \texttt{[CLS]} token to predict the relationship between the two sentences.
	The NSP objective was designed to improve the performance of downstream tasks, such as natural language inference \cite{Bowman,NLI}.
	\subsection{Context-Aware Lexical Substitution}\label{sec:LS}
	\subsubsection{Candidate Set Generation.}
	To generate substitute candidates for the target word $t_i$ in a given sentence $S = \{t_1,...,t_{i-1},t_i,t_{i+1},...,t_N\}$, we first mask the token $t_i$ to get the masked sentence $S_{\backslash i}$, which loses the semantic information carried by $t_i$.
	Motivated by \cite{LSim}, we further concatenate $S$ and $S_{\backslash i}$ with the separator token \texttt{[SEP]} to form 
	\begin{linenomath}
	\begin{equation}
	I_i = Concatenate(S,\texttt{[SEP]},S_{\backslash i}).
	\end{equation}
	\end{linenomath}
	Since $I_i$ contains the complete semantic information of $S$, we feed it into BERT to predict the vocabulary probability distribution of the masked token. Then, excluding the morphological derivations of $t_i$, we choose the top $K$ words as the initial candidate set $W=\{w_1,w_2,...,w_K\}$.
	
	\subsubsection{Inference-Based Candidate Set Ranking.}
	Word rankings in $W$ are still determined by the predicted probability from BERT, which mainly considers the semantic similarity. But it is more important to consider whether the new sentence using the candidate in $W$ can still maintain the same meaning of the original sentence, i.e., the semantic relatedness. As BERT has already demonstrated its strong ability for multi-genre natural language inference (MNLI) in RoBERTa \cite{RoBERTa}, it is very suitable to be used to measure the semantic relatedness of each candidate with the original sentence. Specifically, for each word $w$ in $W$, we use it to replace the target word $t_i$ in $S$, and get the new sentence $\hat{S}=\{t_1,...,t_{i-1},w,t_{i+1},...,t_N\}$. Then we concatenate $\hat{S}$ and $S$ with \texttt{[SEP]} to form 
	\begin{linenomath}
	\begin{equation}
	I^{\prime}_i = Concatenate(S,\texttt{[SEP]},\hat{S}),
	\end{equation}
	\end{linenomath}
	and feed it into the RoBERTa model fine-tuned for MNLI task to inference the relationship (i.e., entailment / contradiction / neutral) between $S$ and $\hat{S}$. Because the probability of `entailment' can indicate the relatedness of two sentences, we propose to use it as the semantic relatedness (SR) measurement to score each candidate. We shall point out that the original sentence $S$ is needed as the reference when calculating the SR score of a candidate. Then we rank the candidates according to their SR scores and get the ranked candidates $RW = \{w^{\prime}_1,w^{\prime}_2,...,w^{\prime}_K\}$. The pseudo code of our LS approach is illustrated in Algorithm \ref{alg:LS}. 
	
	\begin{algorithm}[t]
		\small
		\caption{Context-Aware Lexical Substitution}
		\label{alg:LS}
		\begin{algorithmic}[1]
			\renewcommand{\algorithmicrequire}{\textbf{Input:}}
			\REQUIRE{original sentence $S=\{t_1, t_2,..., t_N\}$, the masked sentence $S_{\backslash i} = \{t_1,...,t_{i-1},\texttt{[MASK]},t_{i+1},..., t_N\}$, candidates generation model $\textsc{Bert}_{gen}$, semantic relatedness scoring model $\textsc{Bert}_{score}$.}
			\renewcommand{\algorithmicensure}{\textbf{Output:}}
			\ENSURE{ranked substitute candidates for $S_{\backslash i}$}
			\STATE $I_i \leftarrow Concatenate(S,\texttt{[SEP]},S_{\backslash i})$
			\STATE $//$ Generate candidates $W$ based on the vocabulary probability  distribution
			\STATE $W \leftarrow \textsc{Bert}_{gen}(I_i)$
			\FOR{each word $w$ in $W$}
			\STATE $\hat{S} \leftarrow \{t_1,...,t_{i-1},w,t_{i+1},...,t_N\}$
			\STATE $//$ Calculate the semantic relatedness score of $\hat{S}$ with $S$ as the reference
			\STATE $I^{\prime}_i \leftarrow {\rm Concatenate}(S,\texttt{[SEP]},\hat{S})$
			\STATE $SR\_score_w \leftarrow \textsc{Bert}_{score}(I^{\prime}_i)$
			\ENDFOR
			\STATE Create the new candidate set $RW$ with all words $w \in W$ ranked by the descending order their SR score
			\RETURN $RW$
		\end{algorithmic}
	\end{algorithm}
	
	\begin{figure*}[t]
		\begin{center}
			\subfigure[Watermarking process.]{
				\includegraphics[width=0.32\textwidth, height=0.35\textwidth]{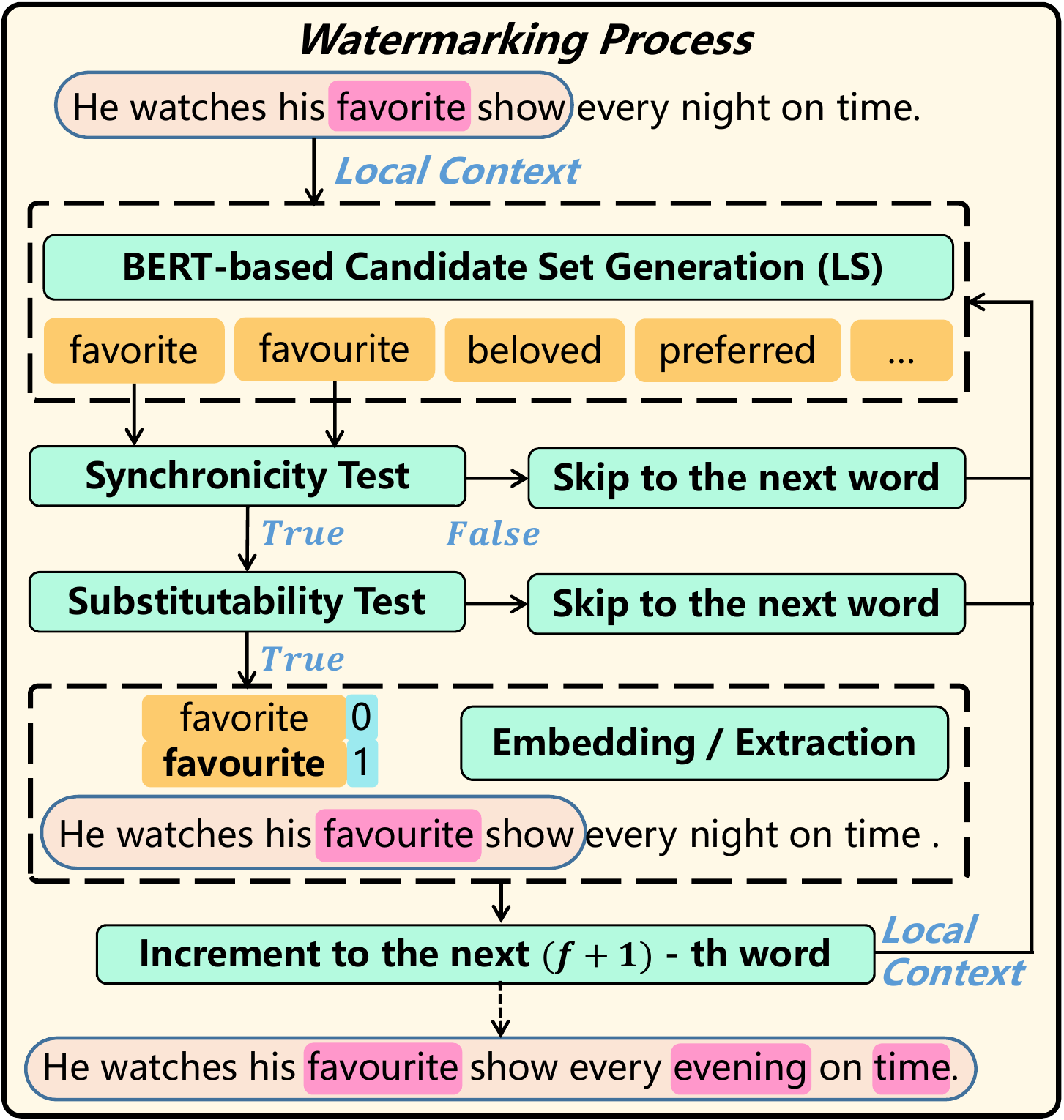}
			}
			\subfigure[Example of embedding.]{
				\includegraphics[width=0.32\textwidth, height=0.35\textwidth]{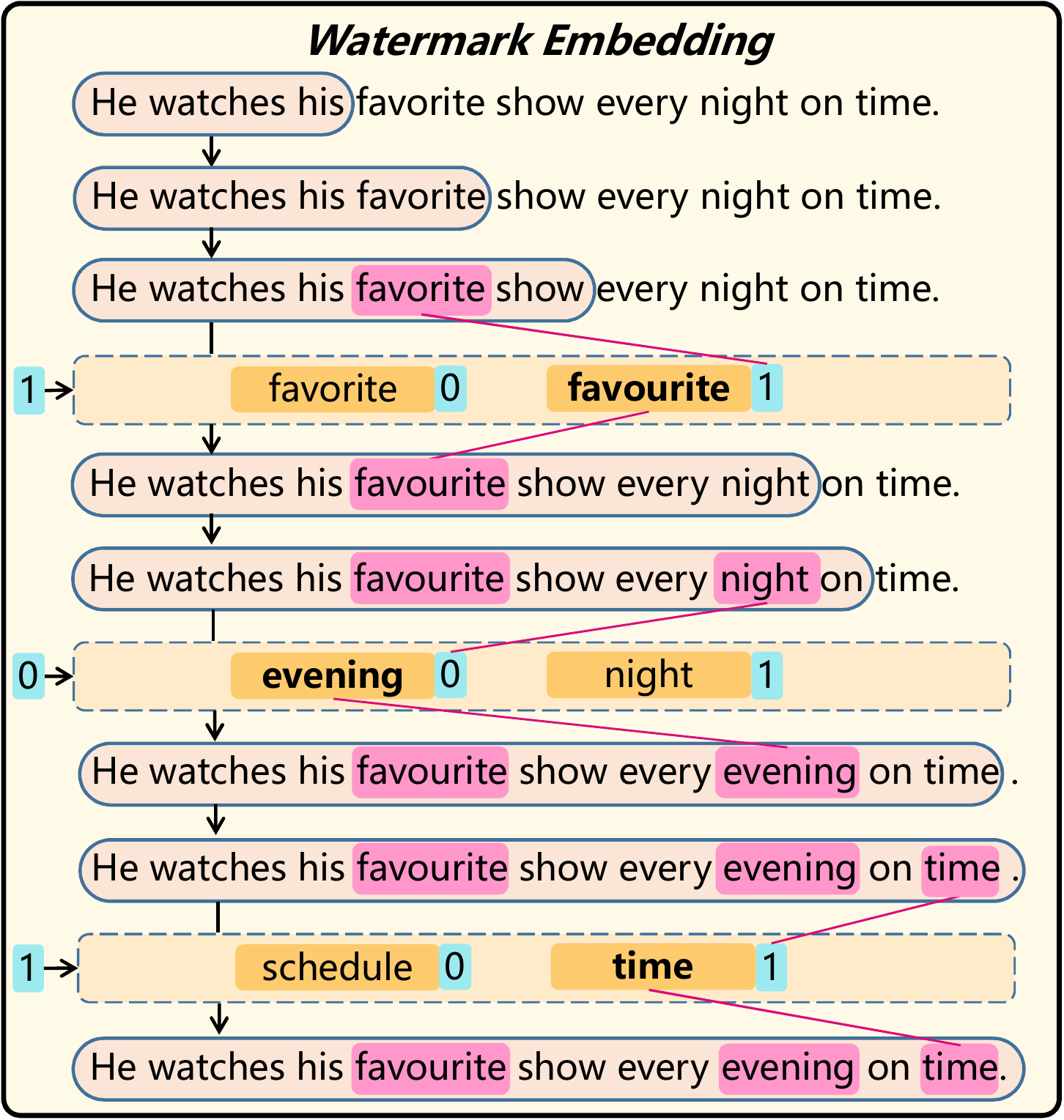}
			}
			\subfigure[Example of extraction.]{
				\includegraphics[width=0.32\textwidth, height=0.35\textwidth]{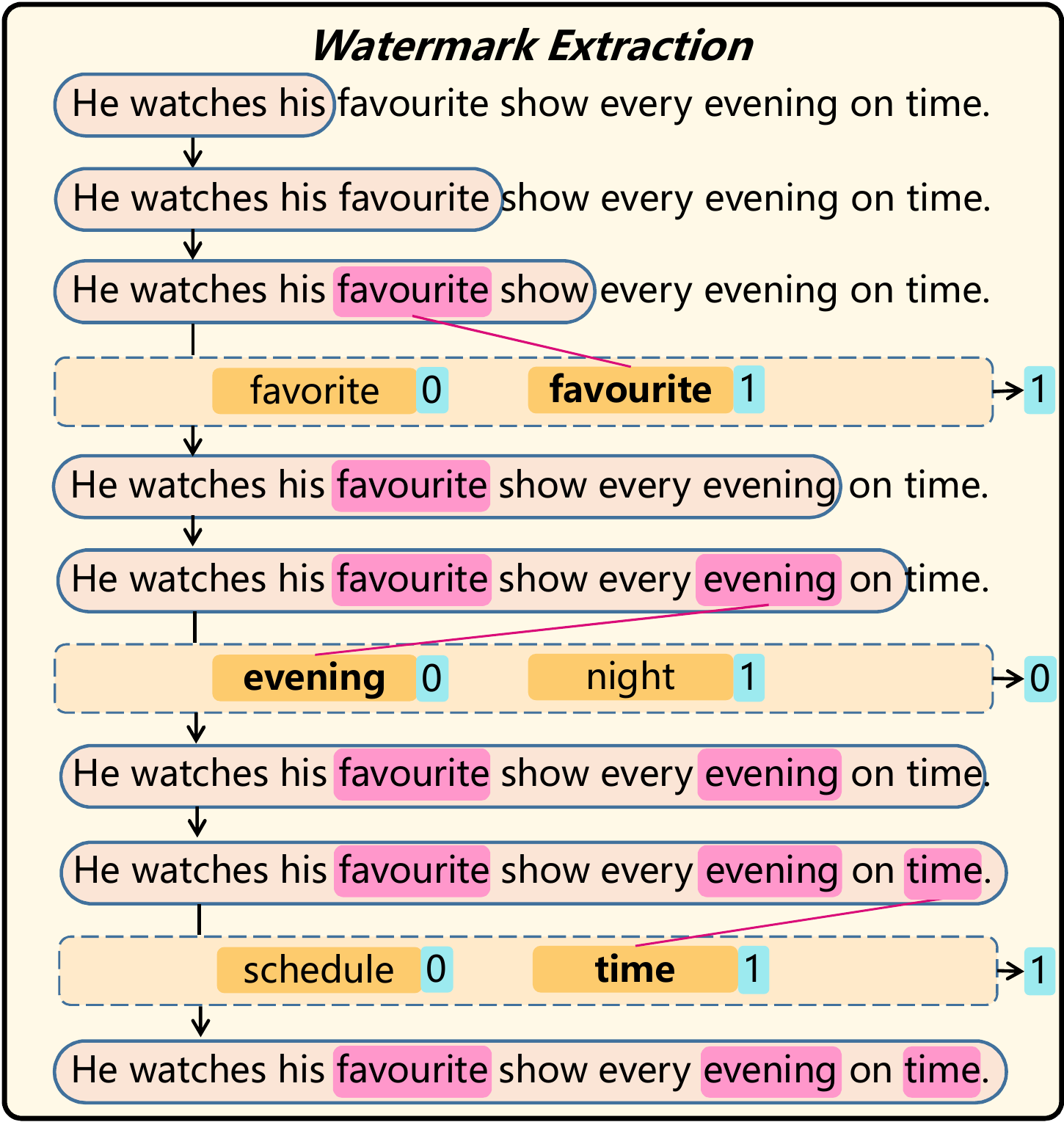}
				\label{fig:framework_example_extract}
			}
			
			\caption{The watermarking process with a step-by-step example. Given the input sentence, we use the synchronicity and substitutability tests to incrementally search and substitute the words capable of carrying watermark signals in the local context.}\label{fig:NLW_framework}
		\end{center}
	\end{figure*}
	
	\begin{table}[t]
		\newcolumntype{L}{>{\arraybackslash}m{5cm}}
		\newcolumntype{C}{>{\centering\arraybackslash}m{1.5cm}}
		\renewcommand{\arraystretch}{1.3}
		\centering
    	\resizebox{1\linewidth}{!}{%
    		\begin{tabular}{L|L|C}
    			\toprule
    			\bf Original Sentence
    			& \bf Substituted Sentence
    			& Candidates identical?
    			\\ \midrule 
    			He watches his \textbf{\underline{favorite}} show every night on time.
    			& He watches his \textbf{\underline{{beloved}}} show every evening on time.
    			& 
    			\\ 
    			\{favorite, \hlc[light_green]{beloved}, favored...\}
    			& \{beloved, favored, loved...\}
    			& \textcolor[rgb]{0.97, 0, 0}{\XSolidBrush}
    			\\ 
    			\centering $\downarrow$
    			& \centering $\downarrow$
    			&
    			\\
    			He watches his beloved show every \textbf{\underline{night}} on time.
    			& He watches his beloved show every \textbf{\underline{evening}} on time.
    			& 
    			\\
    			\{night, \hlc[light_green]{evening}, time, show...\}
    			& \{evening, night, time, moment...\}
    			& \textcolor[rgb]{0.97, 0, 0}{\XSolidBrush}
    			\\\bottomrule
		\end{tabular}}
		\caption{The original sentence and the substituted sentence will generate different candidates for the underlined words with the context-aware lexical substitution approach. Words marked in green are selected to replace the original words.
		}\label{tab:LS_syc_issue}
	\end{table}
	To build our watermarking scheme on the proposed LS approach, there exists a challenge to be solved. Specifically, in the watermark extraction stage, since we only have $\hat{S}$ rather than $S$ and BERT is sensitive to contextual changes, the obtained LS candidates will be different from those generated in the watermark embedding stage, resulting in the extraction failure. 
    An example is shown in Table \ref{tab:LS_syc_issue}, which indicates that it is necessary to synchronize the LS candidates generated in watermark embedding and extraction sides. 
	\subsection{Sequence Incremental Watermarking Scheme}\label{sec:NLW} 
	To solve the challenge mentioned above, we further design the synchronicity and substitutability tests to force the embedding and extraction sides to locate the same words and generate identical candidate sets. Based on it, the sequence incremental watermarking scheme is proposed. One corresponding step-by-step example is illustrated in Figure \ref{fig:NLW_framework}. Before diving into the watermarking process, we first design the synchronicity test for a word. 
	
	\subsubsection{Synchronicity Test.}  Synchronicity means the candidate set generated from a same masked word in both the watermark embedding and extraction sides are identical, even if the original sentence and the watermarked sentence are partly different. To be specific, given a target word $t_i$ in $S=\{t_1,t_2,...,t_N\}$, we want to embed information by replacing it with a word in its candidate set $RW$ generated by Algorithm \ref{alg:LS}. To keep the original semantics as much as possible, we further select words with SR scores higher than $0.95$ and choose the top 2 words in $RW$ as the final candidates $FC=\{w_1^\prime,w_2^\prime\}$. Then, for each word in $FC$, we use it to replace the target word $t_i$ in $S$ to attain the substituted sentences $\hat{S}_1$ and $\hat{S}_2$ respectively. And we repeat the same operations on $\hat{S}_1$ and $\hat{S}_2$ as we did on $S$ to get the candidate sets $FC_1$ and $FC_2$ corresponding to $w_1^\prime$ and $w_2^\prime$. Finally, if $t_i \in FC$ and $FC_1$ and $FC_2$ satisfy the following condition:
	\begin{linenomath}
	\begin{equation} \label{eq:sortstr}
	Sort(FC_1) = Sort(FC_2) = Sort(FC),
	\end{equation}
	\end{linenomath}
	we say the target word $t_i$ has the synchronicity, where $Sort(\cdot)$ is the function to sort strings in ascending order. 
	We represent the synchronicity testing process as follows:
	\begin{linenomath}
	\begin{equation} \label{eq:stest}
	Sync,C = ST(index,S),
	\end{equation}
	\end{linenomath}
	where $ST(\cdot)$ denotes the test function and the inputs are the target word $index$ with its sentence $S$. It returns the target word's synchronicity $Sync$ ($True$ or $False$) and corresponding sorted final candidate set $C=\{c_1,c_2\}$, i.e., the term $Sort(FC)$ in Eq.(\ref{eq:sortstr}). With this synchronicity test, we can find words that can generate the same candidates at the embedding and extraction sides, which allows the message encoding and decoding.
	
	
	
	
	%
	
	\subsubsection{Watermarking Process.}
	Given a text document, we start by splitting it into a list of sentences with the help of the sentence tokenization tools in NLTK\footnote{https://www.nltk.org/}. For each sentence in the list, we propose to embed and extract the watermark information with an incremental local context. The process is detailed in Algorithm \ref{alg:NLW}. Specifically, given the $i$-th word ($2\leq i<N$) in sentence $S=\{t_1,t_2,...,t_N\}$, we test its $Synchronicity$ with the local context consisting of the words ahead of it and the next one word, which can be represented by $L=\{t_1,t_2,...,t_{i+1}\}$. Fed with $i$ and $L$, we calculate the $Sync$ and candidate set $C$ of the $i$-th word by Eq.(\ref{eq:stest}). If $Sync$ is $True$ and $t_i \in C$ (to prevent words like proper names from being substituted), we consider $t_i$ substitutable. Otherwise, we skip it and do the same test for its next word. Considering this skip step, it is necessary to further check whether the substitution of $t_i$ will change  the previous substitution status of word $t_{i-1}$ (\textbf{Substitutability Test}), as described in Algorithm \ref{alg:NLW}, step 14-22.
    
	Finally, in watermark embedding, if $t_i$ is substitutable, we replace it to embed one bit watermark signal with the word in $C$ according to the following rule: 
	\begin{linenomath}
	\begin{equation} \label{eq:embed}
	t_i= \begin{cases}c_1, & \text { if } signal = 0, \\ c_2, & \text { if } signal = 1.\end{cases}
	\end{equation}
	\end{linenomath}
	After one bit of signal embedding, we will get a new sentence $S^\prime$. Here, we require the next word  $t_{i+1}$ unchanged to retain the local context. Then the embedding of the next signal starts from the $(f+1)$-th word of $S^\prime$ with the same process above, where $f$ is the hyperparameter that controls the minimum distance between two substitutions in Algorithm \ref{alg:NLW}. In watermark extraction, the input is the watermarked sentences and all steps are exactly similar to the embedding process, as Figure \ref{fig:framework_example_extract} shows. After locating the substituted word, we generate the candidates that are identical to those in the embedding side and extract the watermark signal by the inverse process of Eq.(\ref{eq:embed}).
	
	\begin{algorithm}[t]
		\small
		\caption{Sequence Incremental Watermark Embedding}
		\label{alg:NLW}
		\begin{algorithmic}[1]
			\renewcommand{\algorithmicrequire}{\textbf{Input:}}
			\REQUIRE{original sentence $S=\{t_1, t_2,..., t_N\}$, the hyper-parameter $f$, the watermark binary bit sequence $m$.}
			\renewcommand{\algorithmicensure}{\textbf{Output:}}
			\ENSURE{watermarked sentence $S_w$}
			\STATE $latest\_embed\_index \leftarrow 0$
			\STATE $index \leftarrow 2$
			\STATE $RiskSet \leftarrow \{punctuations,stopwords,subwords\}$
			\STATE $S_w \leftarrow S_o$
			\WHILE {$index < N - f$}
			\STATE $local\_context \leftarrow \{t_1,t_2,...,t_{index+1}\}$ in $S_w$
			\IF{$t_{index}$ is in $RiskSet$}
			\STATE $index \leftarrow index+1$
			\STATE \textbf{Continue}
			\ELSE
			\STATE $Sync,C \leftarrow ST(index,local\_context)$
			\IF{($t_{index} \in C$) and ($Sync=True$)}
			\STATE $Substitutable \leftarrow True$
			\IF{$(index - latest\_embed\_index) != f+1$} 
			\FOR{each candidate $c$ in $C$}
			\STATE $new\_context \leftarrow \{t_1,t_2,...,c,t_{index}\}$
			\STATE $Sync^\prime,C^\prime \leftarrow ST(index-1,new\_context)$
			\IF{($t_{index-1} \in C^\prime$) and ($Sync^\prime=True$)}
			\STATE $Substitutable \leftarrow False$
			\ENDIF
			\ENDFOR
			\ENDIF
			\ELSE
			\STATE $Substitutable \leftarrow False$
			\ENDIF
			\IF{$Substitutable$ is $True$}
			\STATE Fetch one bit $signal$ that has not been embed in $m$
			\STATE Replace $t_{index}$ in $S_w$ with word in $C$ via Eq.(\ref{eq:embed})
			\STATE $latest\_embed\_index \leftarrow index$
			\STATE $index \leftarrow index+f+1$
			\ELSE
			\STATE $index \leftarrow index+1$
			\ENDIF
			\ENDIF
			\ENDWHILE
			
			\RETURN $S_w$
		\end{algorithmic}
	\end{algorithm}

	\section{Experimental Results}
	In this section, we first provide a detailed introduction of the experiment settings. To demonstrate the effectiveness of our methods, we evaluate the proposed lexical substitution and watermarking methods under some objective metrics. Besides, we conduct a human evaluation on the meaning-preserving ability of the watermarked sentences, since the text content is inherently subjective. Finally, some ablation studies are provided to justify the motivation of our design.
	
	\subsection{Experiment Settings}
	\subsubsection{Dataset.} In the watermarking experiments, we choose four datasets with different writing styles, namely, Novels, WikiText-2, IMDB, and AgNews. For Novels, we select \textit{Wuthering Heights}, \textit{Dracula}, and \textit{Pride and Prejudice} from Project Gutenberg\footnote{https://www.gutenberg.org/}.
	For the rest datasets, we select the first 10,000 sentences each from the WikiText-2, IMDB, and AgNews datasets provided by HuggingFace\footnote{https://huggingface.co/datasets}.
	
	\subsubsection{Implementation Details.} We adopt the pre-trained model \textit{bert-base-cased}\footnote{https://huggingface.co/bert-base-cased} as the candidate generation model $\textsc{Bert}_{gen}$ and \textit{roberta-large-mnli}\footnote{https://huggingface.co/roberta-large-mnli} as the score model $\textsc{Bert}_{score}$. We set $f=1$ by default in Algorithm \ref{alg:NLW} and $K=32$ when generating candidates.
	
	\subsubsection{Comparison Systems.} We compare our method with the WordNet-based methods and the transformer-based method. The former \cite{Topkara1,xiang} generate synonym candidates from WordNet to embed watermarks. And the transformer-based method AWT \cite{awt} trains a data hiding network to substitute the unobtrusive words in the given context.
	
	\subsubsection{Metrics.} Unlike in the field of image watermarking, where objective metrics such as PSNR and SSIM are used to evaluate the quality of watermarked images, there is still no uniform metric for evaluating the semantic quality of the watermarked text. Motivated by using the semantic relatedness (SR) score to rank the candidates in Algorithm \ref{alg:LS}, we choose it to measure the semantic relatedness between the watermarked sentences and original sentences. Besides, we also use the pre-trained sentence transformer model \textit{stsb-roberta-base-v2}\footnote{https://www.sbert.net/} in \cite{sbert} to measure the semantic similarity (SS) between the watermarked sentence and original sentence by computing the cosine distance of their sentences' embeddings. 
	
	\subsubsection{LS Benchmark.} To evaluate our LS approach, we choose the Stanford Word Substitution Benchmark (\textsc{Swords)} \cite{swords}, which is the latest LS benchmark with improved data coverage and quality compared with the past benchmarks. It examines the quality and coverage of the substitutes from the LS approach with respect to the substitutes that humans judged as $acceptable$ or $conceivable$.
	
	\subsection{Results and Discussion}
	\begin{table}[t]
	    \scriptsize
		\begin{center}
			\renewcommand{\arraystretch}{0.8}
			\setlength{\tabcolsep}{5mm}{
				\begin{tabular}{lcccc}
					\toprule[1pt]
					\multirow{2}{*}{Method} & \multicolumn{2}{c}  {Lenient} & \multicolumn{2}{c}  {Strict} 	\\
					\cmidrule(r){2-3} \cmidrule(r){4-5}
					\;  & $F^{10}$  & $F_c^{10}$  & $F^{10}$  & $F_c^{10}$  \\
					\midrule
					\textsc{Humans} &51.6 &76.4 &- &-\\
					\textsc{CoInCo} &34.6 &63.3 &- &-\\
					\textsc{Thesaurus} &17.6 &50.2 &- &-\\
					\midrule
					\textsc{Bert-k} & 31.5 & 53.2 & 15.2 & 23.7 \\
					\textsc{Bert-m} & 30.8 & 47.0 & 10.4 & 16.1\\
					\textsc{Bert}-LS & 31.6 & 53.3 & 16.8 & 26.1 \\
					{\bf Proposed(LS)} &{\bf36.7} &{\bf56.1} & \bf{18.3}  & \bf{28.7}\\
					\bottomrule[1pt]
				\end{tabular}
			}
			\caption{Evaluation of the proposed LS approach on the \textsc{Swords} benchmark. The `Lenient' fashion means the generated substitutes which are not in \textsc{Swords} are filtered out , and `Strict' means the setup without filtering.}
			\label{tab:LS}
		\end{center}
	\end{table}
	
	\subsubsection{Performance on Lexical Substitution.} 
	\begin{table}[ht]
		\begin{center}
			\renewcommand{\arraystretch}{1.1}
			\scalebox{0.62}{
				\begin{tabular}{llcccccc}
					\toprule[1.5pt]
					\multirow{3}{*}{Metric} & \multirow{3}{*}{Method} & \multicolumn{3}{c}{Novels} & \multirow{3}{*}{WikiText-2} & \multirow{3}{*}{IMDB} & \multirow{3}{*}{AgNews}  \\  \cline{3-5}
					&                   & \begin{tabular}[c]{@{}c@{}}\text{Wuthering}\\ \text{Heights}\end{tabular} & \text{Dracula}       & \begin{tabular}[c]{@{}c@{}}\text{Pride} \text{and} \\ \text{Prejudice}\end{tabular} &                 &          &       \\ \midrule
					\multirow{4}{*}{SR}    & Topkara           & 0.8816            & 0.8691          & 0.8956        &    0.8883     & 0.8433          & 0.8587       \\ 
					& Hao             & 0.8930            & 0.9146          & 0.9079      &      0.9072      & 0.8668          & 0.8752      \\ 
					& AWT & 0.9470 & 0.8688 & 0.8897 & 0.9354  & 0.9575&0.9636 \\
					& \textbf{Proposed} & \textbf{0.9844}   & \textbf{0.9852} & \textbf{0.9854}     & \textbf{0.9864} &  \textbf{0.9850} & \textbf{0.9763} \\ \midrule
					\multirow{4}{*}{SS} & Topkara           & 0.9291            & 0.9095          & 0.9314     &     0.9415      & 0.9160          & 0.9694       \\  
					& Hao             & 0.9337            & 0.8886          & 0.9356      &   0.9448    & 0.9426          & 0.9712       \\ 
					& AWT & 0.9677 & 0.8546 & 0.9317 & \textbf{0.9907}  & 0.9727 & 0.9889 \\
					& \textbf{Proposed} & \textbf{0.9888}   & \textbf{0.9861} & \textbf{0.9866} & 0.9892  & \textbf{0.9819} & \textbf{0.9921}  \\ \bottomrule[1.5pt]
				\end{tabular}
			}
			\caption{Evaluation of the semantic relatedness (SR) and semantic similarity (SS) between the original sentences and watermarked sentences of different watermarking methods.}
			\label{tab:NLW_obj}
		\end{center}
	\end{table}
	\begin{table}[ht]
		\newcolumntype{L}{>{\arraybackslash}m{3.3cm}}
		\renewcommand{\arraystretch}{1.5}
		\centering
		\resizebox{1\linewidth}{!}{%
			\begin{tabular}{L|L|L}
				\toprule[1.3pt]
				\textbf{Original} & \textbf{AWT} & \textbf{Proposed} \\\midrule
				resulting in a population decline as workers left for other areas & resulting in a population decline \underline{\textit{\hlc[light_red]{an}}} workers left for other areas & resulting in a \underline{\textit{\hlc[light_green]{demographic}}} decline as \underline{\textit{\hlc[light_green]{employees}}} left for other areas \\
				, but the complex is broken up by the heat of cooking & , \underline{\textit{\hlc[light_red]{and}}} the complex is broken up by the heat of cooking & , but the complex is broken up by the \underline{\textit{\hlc[light_green]{temperature}}} of cooking \\
				Blythe , who is $<$unk$>$ , took off his glasses before entering the stage , which together with the smoke and light effects allegedly left him half & Blythe , who is $<$unk$>$ , took off his glasses before entering the stage , which together \underline{\textit{\hlc[light_red]{@-@}}} the smoke and light effects allegedly left him half  & Blythe , who is $<$unk$>$ , took off his glasses before entering the stage , which \underline{\textit{\hlc[light_green]{along}}} with the smoke and light effects allegedly left him half 
				\\\bottomrule[1.3pt]
		\end{tabular}}
		\caption{Examples of watermarked sentences compared with AWT on WikiText-2. The substituted words are underlined. } \label{tab:awt_compare}
	\end{table}
	We directly evaluate our LS approach on the Stanford Word Substitution Benchmark (\textsc{Swords}). It computes precision $P^k$ and recall $R^k$ at $k=10$, which is
	\begin{linenomath}
	\begin{align}
	P^{k}=\frac{\# \text {\rm\ $acceptable$\ substitutes\ in\ system \ top- } k}{\# \text {\rm\ substitutes\ in\ system\ top-} k},
	\\
	R^{k}=\frac{\# \text {\rm\ $acceptable$\ substitutes\ in\ system\ top- } k}{\min (k, \# \text {\rm\ $acceptable$\ substitutes})}.
	\end{align}
	\end{linenomath}
	Then the harmonic mean of $P_k$ and $R_k$, represented by $F^k$, is calculated. Likewise, it computes $P_k^c$, $R_k^c$, and $F^k_c$ corresponding to the list of substitutes which humans judged as $conceivable$, which is a larger candidate list. For comparison, the sentences with target word either masked (\textsc{Bert-m}) or kept intact (\textsc{Bert-k}) are feed into BERT, and output the top 50 words. \textsc{CoInCo} \cite{coinco} and \textsc{Thesaurus} are the human-crafted candidate sources. As Table \ref{tab:LS} shows, our approach outperforms the state-of-the-art approach (i.e., \textsc{Bert}-LS) in both `lenient' and `strict' setup, which means that our proposed SR score is helpful for BERT to propose LS candidates. 
	\subsubsection{Preserving the Semantics of Original Text.} Using the defined metrics SR and SS, we evaluate the meaning-preserving ability of our watermarking scheme on the datasets with different writing styles. In Table \ref{tab:NLW_obj}, it can be seen that our scheme can well preserve the semantic integrity of the original sentences compared with other natural language watermarking methods. Furthermore, our scheme has good transferability on different datasets, while AWT requires retraining for each dataset. AWT achieves a high SS score on WikiText-2, which is because the sentence embedding is insensitive to the changes of unobtrusive words. But these changes may make the logic and semantics near the changed words incoherent, as shown in Table \ref{tab:awt_compare}. We also used an online automatic grammar checking tool to check the sentences watermarked by AWT and found that some substituted words are easily detected for grammatical errors, which can be found in Appendix C. 
	
	\subsubsection{Human Evaluation.} 
	We randomly sampled 8 sentences on each dataset, marked the substituted words, and asked 10 annotators to rate the effectiveness of the watermarked sentences in maintaining the original meaning with reference to the original sentences. The score ranges from 1 to 5 (very poor to excellent). As Table \ref{tab:userstudy} shows, our method achieves the best performance for preserving the meaning of the original sentences, indicating that our watermarking scheme is more feasible in practical scenarios. We also found that although AWT embed watermarks in the unobtrusive words, such changes were actually abrupt if the original sentence was used as a reference. Some examples used in the evaluation are provided in Appendix A.
	\begin{table}[t] 
	    \scriptsize
		\renewcommand{\arraystretch}{1.1}
		\begin{center}
		    \resizebox{1\linewidth}{!}{
				\begin{tabular}{ccccc} 
					\toprule[1.2pt]
					Method    & Topkara   & Hao
					& AWT
					&\bf Proposed\\ \midrule
					Score & $2.8\pm1.3$ & $2.4\pm1.0$ & $2.0\pm1.2$ & $\bf{4.5 \pm0.6}$  \\ 
					\bottomrule[1.2pt]
				\end{tabular}
			}
		\end{center}
		\caption{The results of human evaluation. The ratings
			range from 1 to 5 (the higher, the better).}
		\label{tab:userstudy}
	\end{table}
	\begin{table}[b] \LARGE
		\begin{center}
			\scalebox{0.5}{ 
				\begin{tabular}{ccccccc} 
					\toprule[1.7pt]
					& \begin{tabular}[c]{@{}c@{}}\text{Wuthering}\\ \text{Heights}\end{tabular}
					&\text{Dracula}
					& \begin{tabular}[c]{@{}c@{}}\text{Pride} \text{and} \\ \text{Prejudice}\end{tabular}
					& IMDB & AgNews & WikiText-2\\ \midrule
					\begin{tabular}[c]{@{}c@{}}Recover \\ proportion\end{tabular}
					& 80.15\% & 81.93\%   & 80.76\%  &82.06\%& 85.25\% & 86.71\% \\ 
					\bottomrule[1.7pt]
				\end{tabular}
			}
		\end{center}
		\caption{The proportion of the substituted words that can be recovered after watermark extraction in different datasets.}
		\label{tab:reconstruct}
	\end{table}	
	\begin{table}[t] \LARGE
		\begin{center}
			\scalebox{0.52}{ 
				\begin{tabular}{ccccccc} 
					\toprule[1.7pt]
					& \begin{tabular}[c]{@{}c@{}}\text{Wuthering}\\ \text{Heights}\end{tabular}
					&\text{Dracula}
					& \begin{tabular}[c]{@{}c@{}}\text{Pride} \text{and} \\ \text{Prejudice}\end{tabular}
					& IMDB & AgNews & WikiText-2\\ \midrule
					\begin{tabular}[c]{@{}c@{}}Payload\\ (bpw)\end{tabular} & 0.081 & 0.090 & 0.080 & 0.097 & 0.088 & 0.105 \\ 
					\bottomrule[1.7pt]
				\end{tabular}
			}
		\end{center}
		\caption{The payload of our watermarking scheme on different datasets, which indicates the average number of bits a word can carry in the datasets.}
		\label{tab:capacity}
	\end{table}
		\begin{table}[t]
		\newcolumntype{L}{>{\arraybackslash}m{5cm}}
		\renewcommand{\arraystretch}{1.2}
		\centering
		\resizebox{1\linewidth}{!}{%
			\begin{tabular}{L|L}
				\toprule[1.3pt]
				\textbf{Embedding Side} & \textbf{Extraction Side}
				\\ \midrule
				In order to achieve this , the \underline{\textit{cooperative}} elements incorporated into the second game were \underline{\textit{removed}} , as they \underline{\textit{took}} up a large portion of \underline{\textit{memory}} space \underline{\textit{needed}} for the improvements .
				&In order to achieve this , the \hlc[light_red]{group} elements incorporated into the \underline{\textit{\hlc[light_red]{subsequent}}} game were \underline{\textit{\hlc[light_red]{omitted}}} , as they \hlc[light_red]{taken} up a large portion of \underline{\textit{\hlc[light_red]{spare}}} space \underline{\textit{\hlc[light_green]{needed}}} for the improvements .
				\\ \midrule
				cooperative - \{cooperative, group\} & -
				\\ 
				second - \{second, subsequent\} & subsequent - \{next, subsequent\} \\
				removed - \{omitted, removed\} & omitted - \{excluded, omitted\} \\
				took - \{taken, took\} & - \\
				memory - \{memory, spare\} & spare - \{save, spare\} \\
				needed - \{needed, required\} & needed - \{needed, required\}
				\\\bottomrule[1.3pt]
		\end{tabular}}
		\caption{A failure case without the \textbf{Synchronicity Test}, where words marked in red indicate they cannot be located (not underlined) or the generated candidate sets are different from the embedding side (underlined).}\label{tab:no_synctest}
	\end{table}
	\subsubsection{Text Recoverability.} According to the synchronicity testing process, the original word must exist in the generated candidate set. Therefore, we try to reconstruct the original text from the watermarked text. Specifically, for each candidate in the candidate set, we mask it and use BERT to predict its probability. Then we rank the two candidates with their probability and choose the top one as the recovered word to replace the corresponding watermarked word to attempt to reconstruct the original sentence. As Table \ref{tab:reconstruct} shows, we find that about $80\%$ of the replaced words in the watermarked sentences can be successfully recovered, which can be used after extracting the watermark message to further preserve the semantics of original sentences. This is also an indication that our method is effective in preserving the semantics of original sentences. See Appendix B for a practical example, where we watermarked the abstract of this paper and recovered some substituted words after watermark extraction.
	\begin{table}[t]
		\newcolumntype{L}{>{\arraybackslash}m{3cm}}
		\renewcommand{\arraystretch}{1.2}
		\centering
		\resizebox{1\linewidth}{!}{%
			\begin{tabular}{cLL}
				\toprule[1.3pt]
				\textbf{Original} & I heard , also , the fir bough repeat its teasing sound , &
				`` I ’ ll put my trash away , because you can make me
				\\\midrule
				\begin{tabular}[c]{@{}c@{}}\text{Embedding Side}\\ \text{(w/ Substitutability Test)}\end{tabular} &I \textit{\hlc[light_green]{heard}} , also , the fir bough repeat its teasing sound , & `` I ’ ll \textit{\hlc[light_green]{place}} my trash away , because you can make me
				\\
				\begin{tabular}[c]{@{}c@{}}\text{Extraction Side}\\ \text{(w/ Substitutability Test)}\end{tabular} &I \textit{\hlc[light_green]{heard}} , also , the fir bough repeat its teasing sound , & `` I ’ ll \textit{\hlc[light_green]{place}} my trash away , because you can make me
				\\ \midrule
				\begin{tabular}[c]{@{}c@{}}\text{Embedding Side}\\ \text{(w/o Substitutability Test)}\end{tabular} & I \textit{\hlc[light_green]{heard}} , also , the fir bough repeat its teasing \textit{\hlc[light_red]{noise}} , 
				& `` I ’ ll \textit{\hlc[light_green]{place}} my trash \textit{\hlc[light_red]{aside}} , because you can make me 
				\\
				\begin{tabular}[c]{@{}c@{}}\text{Extraction Side}\\ \text{(w/o Substitutability Test)}\end{tabular} & I \textit{\hlc[light_green]{heard}} , also , the fir bough repeat its \textit{\hlc[light_red]{teasing}} noise , & `` I ’ ll \textit{\hlc[light_green]{place}} my \textit{\hlc[light_red]{trash}} aside , because you can make me 
				\\\bottomrule[1.3pt]
		\end{tabular}}
		\caption{Comparison of word locating results with and without the \textbf{Substitutability Test}, where words marked in light green indicate successful locations and words in light red indicate failed locations.}\label{tab:no_subtest}
	\end{table}
	\begin{table}[t] 
     \scriptsize
		\renewcommand{\arraystretch}{0.8}
		\begin{center}
			\setlength{\tabcolsep}{6.6mm}{
				\begin{tabular}{cccc}
					\toprule[1.3pt]
					$f$            & 1      & 2      & 3      \\ \midrule
					SR        & 0.983 & 0.984 & 0.985 \\ 
					SS & 0.988 & 0.994 & 0.995 \\ \midrule
					Payload (bpw)& 0.091 & 0.044 & 0.031 \\ \bottomrule[1.3pt]
			\end{tabular}}
		\end{center}
		\caption{The average semantic quality score and payload with different values of $f$.}
		\label{tab:f123}
	\end{table}
	\subsubsection{Payload and Robustness.} In Table \ref{tab:capacity}, we show the average payload of our watermarking scheme on the different datasets. The payload is the average amount of information that one single word can carry, and is in unit of \textit{bits per word (bpw)}. For the robustness, due to the watermark embedding in semantic dimension, our watermarking scheme are naturally robust to cross-media attacking such as print-camera, screen-camera shooting, print-scanning, OCR, retyping, etc. So the leakage source or plagiarist of the watermarked illegal copies in these scenarios can be traced by extracting the watermark information with a $0\%$ bit error rate (BER). 
	\subsection{Ablation Study}
	\subsubsection{The Importance of Synchronicity Test.} The purpose of the synchronicity test is to ensure that the candidate sets obtained on the extraction side are identical to the ones generated on the embedding side, based on the located word. As shown in Table \ref{tab:no_synctest}, the watermark extraction fails if there is no synchronicity test. Specifically, it fails to locate the watermarked words (\eg 'group' and 'took') or the generated candidates are different from the embedding side (\eg 'removed' vs 'omitted'). Moreover, without this constraint, some special words that are not suitable to be modified may be replaced (\eg the proper noun: `memory').
	\subsubsection{The Importance of Substitutability Test.} We show in Table \ref{tab:no_subtest} the synchronization failures caused by not performing the substitutability test. This is because substituting a word may change the status of its previous word from non-substitutable to substitutable, so that the words located at the extraction side may be different from the embedding side.
	\subsubsection{The Influence of Different Values of $f$.} To discuss the impact of different values of $f$ in Algorithm \ref{alg:NLW}, we set $f=1, 2, 3$ to evaluate the semantic quality and payload of the watermarked sentences. As Table \ref{tab:f123} shows, the average payload decreases rapidly when $f$ grows, but the semantic score will not change significantly. So we set $f=1$ by default to achieve a larger capacity.
	
	\section{Conclusion} 
	In this paper, we first introduce the inference-based semantic relatedness into lexical substitution and leverage it to propose a new context-aware LS approach. Further, to embed watermarks with the proposed LS approach, we design the synchronicity and substitutability tests to locate the words capable of carrying watermark signals in the local context. Compared with existing methods, the proposed watermarking scheme can well preserve the semantics of original sentences and has a better transferability across different writing styles. We hope this paper can attract more attention in this field and inspire more great works.
	
	\section*{Acknowledgements}
	
	This work was supported in part by the Natural Science Foundation of China under Grant 62072421, 62102386, 62002334, 62121002, and U20B2047, Anhui Science Foundation of China under Grant 2008085QF296, Exploration Fund Project of University of Science and Technology of China under Grant YD3480002001, and by Fundamental Research Funds for the Central Universities under Grant WK2100000011 and WK5290000001.

	\bibliography{reference}
	
\end{document}